\documentclass[twocolumn, pra, showpacs,superscriptaddress]{revtex4}%
\usepackage{graphicx}
\usepackage{amsmath}
\usepackage{bm}
\usepackage{float}
\usepackage{color}
\usepackage{sidecap}
\usepackage{soul}
\usepackage[colorlinks, citecolor=blue]{hyperref}
\usepackage{amsfonts}
\usepackage{amssymb}%
\providecommand{\U}[1]{\protect\rule{.1in}{.1in}}
\setstcolor{red}
\begin{document}

\title{Electromagnetically Induced Transparency in a Buffer Gas Cell with Magnetic Field}
\author{Hong Cheng}
\affiliation{State Key Laboratory of Magnetic Resonance and Atomic and Molecular Physics,
Wuhan Institute of Physics and Mathematics, Chinese Academy of Sciences, Wuhan
430071, People's Republic of China}
\affiliation{University of Chinese Academy of Sciences, Beijing 100049, People's Republic
of China}
\author{Han-Mu Wang }
\affiliation{State Key Laboratory of Magnetic Resonance and Atomic and Molecular Physics,
Wuhan Institute of Physics and Mathematics, Chinese Academy of Sciences, Wuhan
430071, People's Republic of China}
\affiliation{University of Chinese Academy of Sciences, Beijing 100049, People's Republic
of China}
\author{Shan-Shan Zhang }
\affiliation{State Key Laboratory of Magnetic Resonance and Atomic and Molecular Physics,
Wuhan Institute of Physics and Mathematics, Chinese Academy of Sciences, Wuhan
430071, People's Republic of China}
\affiliation{University of Chinese Academy of Sciences, Beijing 100049, People's Republic
of China}
\author{Pei-Pei Xin }
\affiliation{State Key Laboratory of Magnetic Resonance and Atomic and Molecular Physics,
Wuhan Institute of Physics and Mathematics, Chinese Academy of Sciences, Wuhan
430071, People's Republic of China}
\affiliation{University of Chinese Academy of Sciences, Beijing 100049, People's Republic
of China}
\author{Jun Luo}
\affiliation{State Key Laboratory of Magnetic Resonance and Atomic and Molecular Physics,
Wuhan Institute of Physics and Mathematics, Chinese Academy of Sciences, Wuhan
430071, People's Republic of China}
\affiliation{University of Chinese Academy of Sciences, Beijing 100049, People's Republic
of China}
\author{Hong-Ping Liu \footnote{Email:liuhongping@wipm.ac.cn}}
\affiliation{State Key Laboratory of Magnetic Resonance and Atomic and Molecular Physics,
Wuhan Institute of Physics and Mathematics, Chinese Academy of Sciences, Wuhan
430071, People's Republic of China}
\affiliation{University of Chinese Academy of Sciences, Beijing 100049, People's Republic
of China}

\begin{abstract}
We have studied the phenomenon of electromagnetically induced transparency
(EIT) of $^{87}$Rb vapor with a buffer gas in a magnetic field at room temperature. It is found that the spectral lines caused by the velocity selective optical pump effects get much weaker and wider when the sample cell mixed with a 5-Torr N$_2$ gas while the EIT signal kept  almost unchanged.
 A weighted least-square fit is also developed to remove the Doppler broadening completely. This spectral method provides a way to measure the Zeeman splitting with high resolution, for example, the $\Lambda $-type EIT resonance splits into four peaks on the $D_2$ line of $^{87}$Rb in the thermal 2-cm vapor cell with a magnetic field along the electric field of the linearly polarized coupling laser. The high resolution spectrum can be used to lock the laser to a given frequency by tuning the magnetic field.

\end{abstract}
\date{\today }
\maketitle
\volumeyear{ }
\volumenumber{ }
\issuenumber{ }
\eid{ }
\received[Received text]{}
\revised[Revised text]{}
\accepted[Accepted text]{}
\published[Published text]{}
\startpage{1}
\endpage{ }

\section{Introduction}

The interesting coherent interaction of atoms with laser fields in three
level $\Lambda $ configurations has attracted increasing attention in
studies of nonlinear and quantum optics and spectroscopy recently.
Electromagnetically induced transparency (EIT) is a quantum coherence effect
which has a number of important applications such as laser cooling \cite{975},
lasing without inversion \cite{1121}, information storage \cite{976} and
magnetometry \cite{1080}.

EIT resonance line-shape and line-width are of interest for many EIT
applications. A narrow and high contrast EIT resonance is useful for many
fields, especially for the study of spectroscopy \cite{1531,1397} and precision metrology \cite{795,1196}. In
a three-level $\Lambda $ scheme, the line-width caused by the nonlinear
effects in EIT is governed by the relaxation rate of ground state coherence,
which is predominantly determined by the interaction time of the atom with
the applied laser fields, i.e, by the average time-of-flight of an atom
through the laser beam \cite{1162,1089}. To get a narrower EIT resonance, a
buffer gas can be added to the atomic vapor to prolong the interaction time
\cite{1162,1424}. The addition of a buffer gas slows the diffusion of the
coherently prepared atoms through the laser beam by simultaneously
preserving the ground state coherence as a result of the collisions over
millions of buffer-gas \cite{1068}. The line-width can be reduced by several
orders of magnitude \cite{1062,1471,1052}.

The line-shape of EIT resonance is also affected by the thermal motion of
atoms in a vapor cell filled with a mixture of alkali atoms and inert gas at
several Torr \cite{1103,1051,1044,483,1489}. For example, Sarkisyan studied
the effect of different pressure of buffer gas on the EIT resonances \cite{1103} while Eugeniy et al. studied the influence of the laser detuning on
the line shape of the EIT and observed that in the presence of a buffer gas
an absorption resonance appears, which is different to the Lorentzian shape
of the usual EIT resonance \cite{1051,1216}.

What's more, compared to the case without buffer gas, the velocity selective
optical pump (VSOP) effects when the vapor cell filled with a buffer gas
will disappear due to the pumping process overridden by velocity-changing.
Therefore, the EIT resonance will grow up on a broaden Gaussian background,
which  simplifies the analysis of the spectra and supplies a good way for us
to study the EIT splitting with a high resolution spectrum in a magnetic
field \cite{1077}.

The EIT resonance in the $\Lambda $-system of the $D_{1}$ or $D_{2}$ line of
Rb atoms in a magnetic field has been studied a lot both theoretically and
experimentally \cite{1102,504,1451,1203}. In most cases, the magnetic field is
either parallel to the laser propagation direction, where both coupling and
probe fields are seen as combination of left- and right-circularly polarized
ones, or orthogonal to the laser propagation direction but along the
electric field of the probe laser, where only the coupling beam is seen as
combined fields \cite{1498}.

However, for the case with the magnetic field along the electric field of
the coupling laser the Zeeman-splitting in magnetic fields is rarely
studied. In our work, we use a vapor cell filled with $^{87}$Rb and 5 torr N$%
_{2}$ to remove the VSOP peaks in the $\Lambda $-type EIT for $D_{2}$ line
at room temperature and present an experimental observation of $\Lambda $%
-type EIT by applying an external magnetic field along the electric field of
the coupling laser. A skill of weighted least-square fit is used to remove the Doppler broadening background. The observed spectrum is well explained by our theoretical simulation.

\section{Experimental setup}

\begin{figure}[ptb]
\begin{center}
\includegraphics[width=3in]{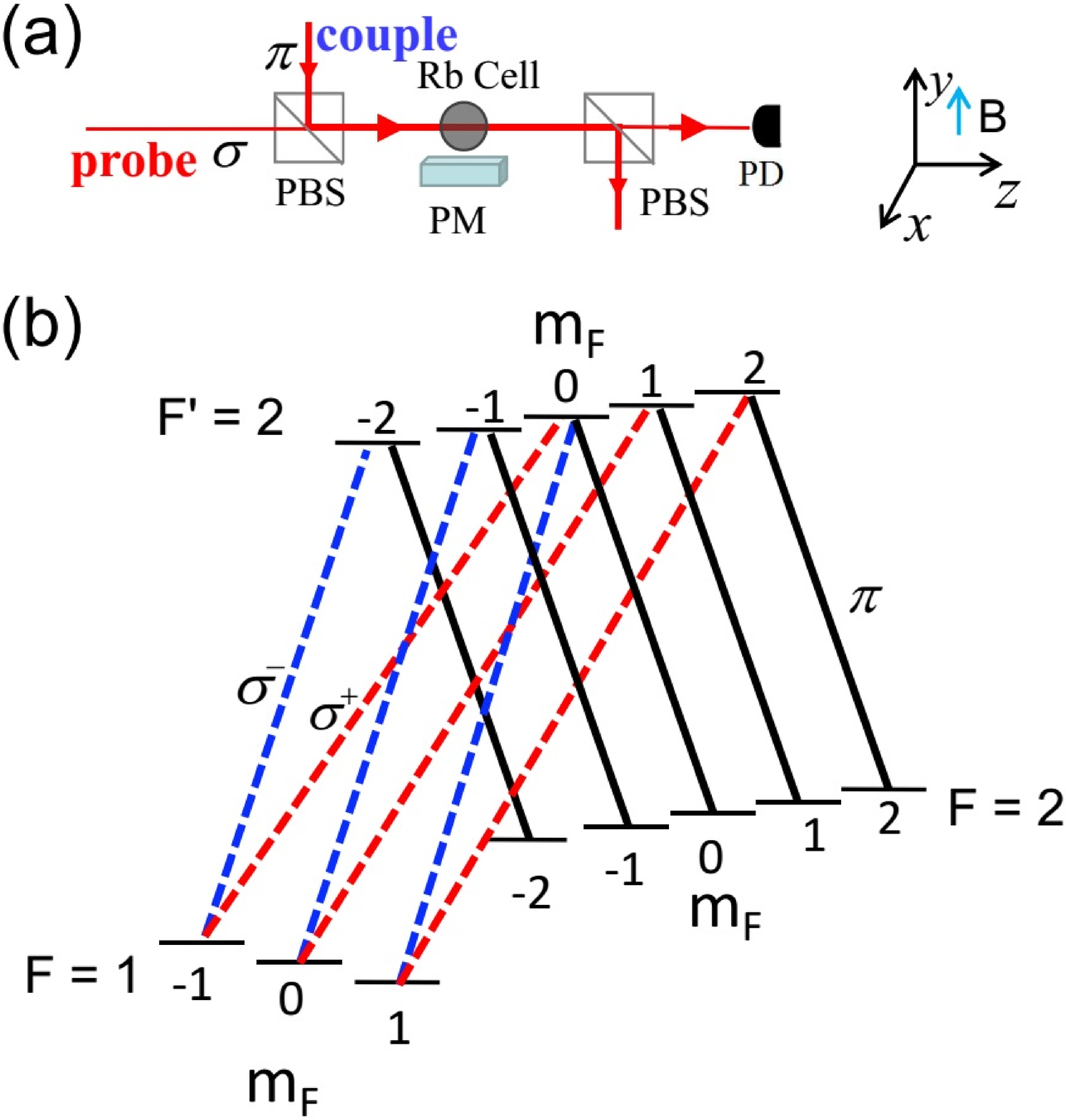}
\end{center}
\caption{(Color online) (a) Experimental setup for EIT spectral measurement.
PBS: polarizing beam splitter; PM: permanent magnet; PD: photo detector. (b)
Magnetic sublevels diagram of $^{87}$Rb atoms in the presence of magnetic
field. All possible polarization combinations for $\Lambda$-EIT scheme are
shown.}
\label{fig1}
\end{figure}

Our experimental setup for the $\Lambda -$type EIT of $^{87}$Rb in magnetic
field is shown in Fig. \ref{fig1}(a) and the relative transition energy
level diagram shown in Fig. \ref{fig1}(b). The arrangement for the optical
elements is also detailed in our previous work \cite{1496}. The coupling
laser frequency is locked to a saturation absorption
spectroscopy (SAS) system and the probe laser is scanned across the $D_{2}$
line levels.

The polarizations of the coupling and probe lasers are linear and mutually
orthogonal. The beam size is around $1$ mm$^{2}$ for each laser. The
coupling and probe laser beams are carefully superimposed on one
polarization beam splitter (PBS) and then adjusted to overlap almost
completely throughout the cell. The spherical shaped Rb vapor cell is made
of pyrex glass and has a size of diameter $2$ cm. The cell is filled with
pure $^{87}$Rb and N$_{2}$ buffer gas at a pressure of $5$ Torr at room
temperature. There is no magnetic field shielding outside the Rb vapor cell.
A long permanent magnet is placed near the cell to apply a magnetic field
for the Rb atoms. The magnetic field direction is along the $y-$axis, which
is parallel to the electric field of the coupling laser, and the strength
can be controlled by changing the distance between the magnet and the cell.
The strongest magnitude can be up to $500$ Gauss. The coupling field couples the
hyperfine level $F=2$ of the ground state ${5}S_{1/2} $ and $F^{\prime }=2$
of the excited state ${5}P_{3/2}$ while the probe laser scans over the
transitions from the ground state $F=1$ to the excited states ${5}P_{3/2}$ $%
F^{\prime }=0,1$ and $2$.

\section{Results and discussion}

In fact, the $D_{2}$ line of $^{87}$Rb atoms are of multi-levels, and there
will be some satellite peaks along with  the EIT line due to the Doppler effect
in the $\Lambda $ configuration \cite{473}, which causes the observed
spectrum very complicated. In our work, we use a pure
$^{87}$Rb vapor cell filled with N$_{2}$ buffer gas to decrease or remove this effect and only
keep the EIT signal for high spectroscopy, which further helps us to study
the EIT splitting in the magnetic field.

\begin{figure}[ptbh]
\begin{center}
\includegraphics[width=3.3in]{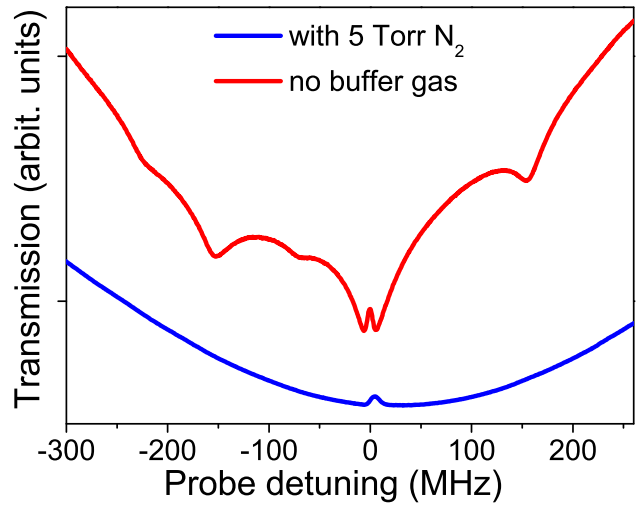}
\end{center}
\caption{(Color online) The experimental observation of the EIT spectra
without and with buffer gas. The coupling laser is locked to the crossover
between $F'=1$ and $2$.}
\label{fig2}
\end{figure}

Figure \ref{fig2} shows the observed EIT spectra without and with N$_{2}$
buffer gas under no magnetic field. The coupling laser is locked to the transition
$F=2$ to $F^{\prime }=1\&2$, where $F^{\prime }=1\&2$ means the cross-over
peak between $F^{\prime }=1$ and $2$ due to the Doppler broadening in SAS
configuration. In the case without buffer gas we can see an EIT peak
with several VSOP peaks. Obviously, the EIT signal is much narrower than
that of VSOP. However, if we fill the cell with 5 Torr N$_{2}$, all the VSOP
peaks will get much broadened and its magnitude also decreases greatly.
This feature keeps the EIT peak to emerge on a very
broadened background and therefore the effects of VSOP will not affect our
observation for the spectral change of atom in external field in the ground
state.

\begin{figure}[tbph]
\begin{center}
\includegraphics[width=3.3in] {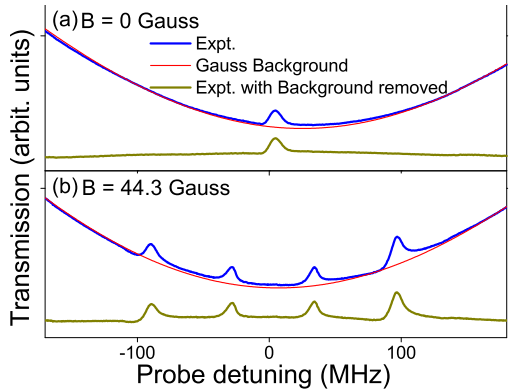}
\end{center}
\caption{(Color online) Experimental spectrum of Rb atom with buffer gas and
its removing of the broadening background. In magnetic field either $B=0$ (a) or $B=44.3$ Gauss (b), we can
get a nice signal to noise ratio, thus  providing a technique to study spectrum
of high resolution. }
\label{fig3}
\end{figure}

To remove the Doppler broadening background, we developed a numerical skill based on the least-square fit.
The main idea is to set the weights of data
points around the EIT peaks to zero and a fit of the data gives a Gaussian
background which we can remove from the experimental data by a subtraction.
Shown in Figs. \ref{fig3}(a)(b) are the experimental spectra  without and with the magnetic field and their numerical processes, where we can see that the Doppler broadening is removed completely.
In Fig. \ref{fig3}(b), the magnetic field is $B=44.3$ Gauss and along the $y-$axis, where the EIT spectral line splits
into four peaks.

\begin{figure}[tbph]
\begin{center}
\includegraphics[width=3.3in] {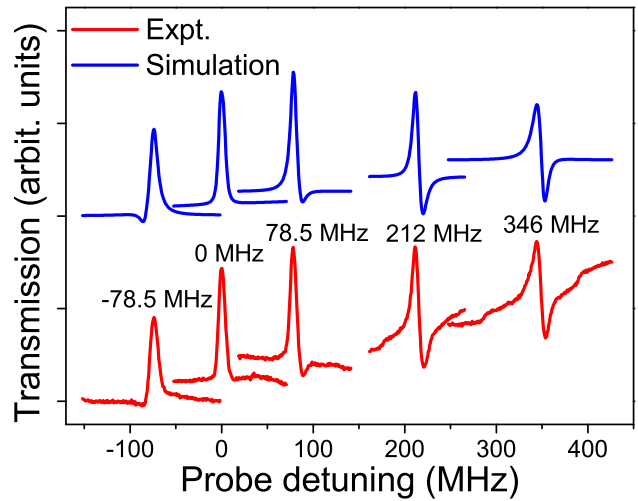}
\end{center}
\caption{(Color online) The effect of detuning of coupling laser on the EIT
spectral line with $B=0$. A large detuning distorts the EIT spectral line
seriously. }
\label{fig4}
\end{figure}

As observed in many previous works \cite{1525,1051,1212,1193}, the EIT spectral line-shape
becomes asymmetric when the coupling laser detuned far away from resonance,
which is re-investigated in our experiment as shown in Fig. \ref{fig4}. The detuning of the
coupling laser varying from $-78.5$ to $346$ MHz causes the EIT resonance
shifting within the absorption profile of the probe laser and the lineshape of the detuned EIT resonance becomes asymmetric and
non-Lorentzian. In addition, an increasing detuning of the coupling frequency leads to  a significant reduction of the amplitude of the EIT peaks \cite{1451}, even inducing a dispersive line shape. For example, when the coupling field frequency detuned to $346$ MHz, an obvious dispersion-like resonant absorption peak appears. Our simulation considering the detuning coincides well with the experimental spectrum.
If we apply a magnetic field to the atom, the hyperfine energy levels will split and
the coupling laser detunes away from the resonant energies as well.
 It also causes the EIT
spectral lines shows an asymmetric structure as  shown in Fig. \ref{fig3}(b).

Unlike the case of atom in zero magnetic field, the degeneracy of the
hyperfine levels is destroyed due to the Zeeman effect,
forming 13 nondegenerate magnetic sublevels as shown in Fig. \ref{fig1}(b).
Specially, for the case where the magnetic field is parallel to the electric
field vector of the coupling laser along $y-$axis, we can simplify
the complex transitions by decomposition. The coupling process
corresponds to a $\pi $ transition while the probe beam has
a $\sigma $ one which can be decomposed to be the combination of  $\sigma^+$ and $\sigma^-$ transitions in the same weight.
All possible transitions can form six $\Lambda-$type subsystems. The two sub-EIT peaks in the outside each contains one single subsystem, corresponding to the transitions from $F=1,m_{F}=1$
via $F^{\prime }=2,m_{F}^{\prime }=2$ to $F=2,m_{F}=2$ and $F=1,m_{F}=-1$
via $F^{\prime }=2,m_{F}^{\prime }=-2$ to $F=2,m_{F}=-2$, respectively. While the two sub-EIT peaks in the middle both contain two subsystems.

\begin{figure}[ptbh]
\begin{center}
\includegraphics[width=3.3in]{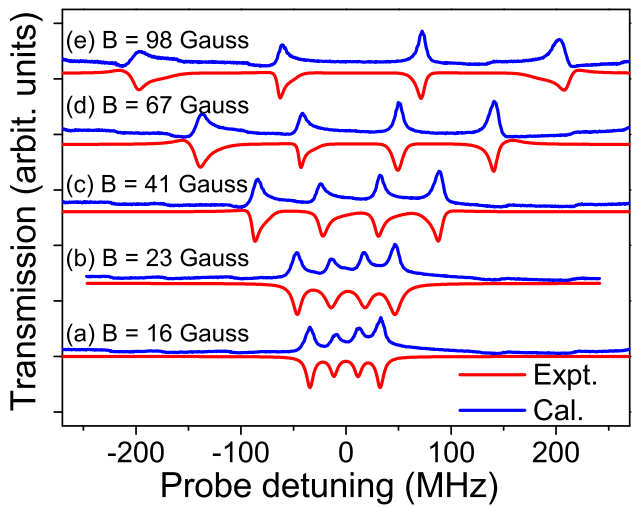}
\end{center}
\caption{(Color online) The observed and calculated splittings of EIT resonances  in magnetic fields along the coupling laser
polarization direction but at different magnetic field intensities. }
\label{fig5}
\end{figure}

Shown in Fig. \ref{fig5} is the spectrum of $^{87}$Rb in magnetic fields along the $y$-axis but at different magnetic field intensities. The magnetic field varies
from $B=16$ Gauss to $B=98$ Gauss. Firstly, we can see that the four EIT peaks are linearly shifted away with the
magnetic field increasing. Secondly, in the low magnetic field ($B=16$ Gauss), the
amplitudes of the middle two peaks are smaller than the  two peaks outside.
An interesting phenomenon is that with the magnetic field
increasing, the amplitudes of the two middle  peaks become larger and larger than the ones outside. They do not keep a constant ratio at all as expected.

Usually the spectral line intensity is determined directly by
the transition probability between different Zeeman sublevels, in our case, from $F=2$ to $F'=2$.
For example, the transition probabilities from $F=2, m_{F}=\pm2$ to $F'=2, m_{F}
=\pm2$ (corresponding to the two outside peaks) are two times larger than that of the transitions $F=2,
m_{F}=\pm1$ to $F'=2, m_{F}=\pm1$ (corresponding to the two inside peaks)\cite{1526}. It agrees with the observation at low magnetic field, i.e., at $B=16$ Gauss.

\begin{figure}[ptbh]
\begin{center}
\includegraphics[width=3.3in]{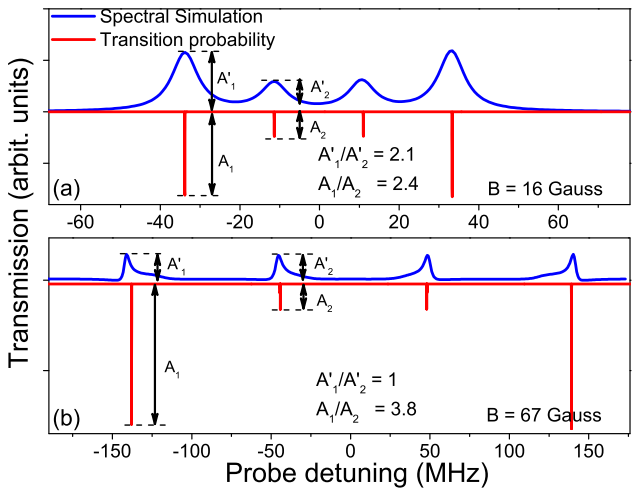}
\end{center}
\caption{(Color online) The results of with or without considering the detuning of the coupling laser in the calculation at different magnetic field intensities. }
\label{fig6}
\end{figure}

However, as discussed previously, the magnitude of the resonance is also dependent on the detuning of the coupling laser frequency from the corresponding atomic transition.
To show the influence of the detuning caused by the magnetic field on the magnitude of the EIT resonance, we have a further comparison of with or without considering the detuning in our calculation, which is  as shown in Fig. \ref{fig6}. The blue and red lines indicate the calculated results of considering or without considering the detuning, respectively.
In the low magnetic field as shown in Fig. \ref{fig6}(a), the simulated spectral line intensity ratio $A_1'/A_2'$ having considered the detuning  is almost the same with the ratio $A_1/A_2$ of the transition intensity, which indicates that in the low magnetic field, the amplitudes of EIT peaks are mainly determined by the transition intensities between different sub-levels and we can ignore the influence of the detuning on the amplitudes.

However, the detuning influence plays a more and more important role for the line intensity and position when the  magnetic field increased. It is shown in Fig. \ref{fig6}(b), where we can see that the amplitude ratio $A_1'/A_2'$ considering the detuning decreased to $1$ from its original ratio $A_1/A_2 =3.8$. This is because the outside EIT peaks own a much larger  red- or blue-detuning, which distorts the resonant lineshape and reduces the amplitudes of their spectral line intensities significantly \cite{1525}. For example, for the magnetic field of $B=98$ Gauss shown in Fig. \ref{fig5}(e), the detunings of the coupling laser for the peaks on the two outsides can reach more than $200$ MHz, resulting in a remarkable reduction of the spectral amplitudes. In addition, the simulated spectral line position also shifts a little. It also indicates that we have to choose a zero detuning line if this Doppler
free technique is used for a  laser frequency locking.

\section{Conclusion}

In this paper, we have observed a high resolution of EIT splittings in a Rb vapor
cell with N$_{2}$ buffer gas in a magnetic field at room temperature. With
the additional 5 Torr N${_{2}}$, the VSOP peaks due to optical pumping
almost disappeared and only the EIT signals were kept in the spectra due
to
the collision effects by the buffer gas.
We present a method by a least-square fit to remove the Doppler broadening
completely which keeps only the EIT peaks. These two techniques simplify
 the analysis of the
observed EIT spectral splittings in the magnetic field. The dependence of splitting line
shape and intervals of sub-EIT windows on the magnetic field have been
investigated. When the applied magnetic field is along the polarization direction
of the coupling laser, the EIT peak splits into four sub-ones. An asymmetric
spectral lineshape can be observed for the deep detunings caused by the
Zeeman level splitting. All experimental observations are well explained
by a simulation considering the detuning effect. Our work provides a method to  lock the laser to a given frequency by tuning the magnetic field.

\end{document}